\documentclass[useAMS,usenatbib]{mn2e}
\usepackage{natbib}
\usepackage{longtable}
\usepackage{rotating}
\usepackage{graphics}
\usepackage{array}
\usepackage{graphicx}
\usepackage{amssymb}
\usepackage{array}
\usepackage{times}
\begin{document}
\title[Photometry of comet C/2007 N3(Lulin)]{
Near opposition photometry of comet C/2007 N3 (Lulin)}
\author[U. C. Joshi, S. Ganesh, K. S. Baliyan]{U.C. Joshi\thanks{e-mail: joshi@prl.res.in},
S. Ganesh, and K. S. Baliyan\\
Astronomy \& Astrophysics Division, Physical Research Laboratory,  Navrangpura, Ahmedabad 380009, India}
\date{Submitted to MNRAS }
\pagerange{\pageref{firstpage}--\pageref{lastpage}} \pubyear{2010}
\maketitle
\label{firstpage}
\begin{abstract}

Observations on comet C/2007 N3 (Lulin) were made at phase angles close to
opposition ($1.7^{\circ}$ - $10^{\circ}$). Photometric observations were carried
out during 2009 February 24-28, in the IHW blue and red continuum and $R$ broad
band using photo-polarimeter mounted on the 1.2m telescope at Mt Abu IR
Observatory.  In all the bands, a significant linear increase in brightness with
decreasing phase angle is detected for the above phase angle range. The phase
coefficient ( $\beta = 0.040\pm 0.001$ mag deg$^{-1}$ estimated in IHW red
(6840\AA) filter band) is found to be independent of wavelength. No non-linear
opposition surge is observed for phase angle $>1.7^{\circ}$. The linear increase
in brightness with decreasing phase angle in the range mentioned earlier can be
explained using the shadow hiding model.  The colour of the comet is found to be
similar to the solar colour indicating the dominance of grains larger than
$0.1\mu m$. A dip in the brightness of about 0.20 mag is seen at the phase angle
$\sim 6.5^\circ$.

\end{abstract}
\section {Introduction}

Comet C/2007 N3 (Lulin) was discovered as an asteroidal object by Quanzhi
Ye on images acquired by Chi Sheng Lin in the course of the Lulin Sky Survey 
and later Young reported it showing marginal cometary appearance on CCD images
taken with the Table Mountain 0.61-m telescope \citep{Ye2007}. 
In February 2009 comet was not only closest to the Earth but it happened to be
close to the opposition. It has been observed that in the  region within a few
degrees of zero phase angle, the reflectance of many Solar system bodies and
also of particulate samples measured in the laboratory increases non-linearly as
the phase angle decreases\citep[see][for a review]{hapke93}. This is known as
"the opposition effect" \citep{gehrels56}. The first attempt to explain the
opposition effect was offered as a consequence of the reduction of mutual
shadows cast between regolith grains as phase angle decreases
\citep[see][]{hapke63, hapke86}. However, this hypothesis fails to explain the
strong opposition effect observed in highly reflecting media
\citep[see][]{harris89, brown83}. An alternative explanation of the opposition
surges in highly reflecting media was offered by \citet{shkuratov85},
\citet{muinonen90}, and \citet{hapke90} invoking the hypothesis of coherent
constructive interference, also called coherent backscattering. \citet{hapke93,
hapke98}  applied it to explain the opposition surge in lunar samples. The
coherent backscattering has also been invoked to explain the reflectance and
backscattering of Saturn's ring \citep{mishchenko92a, mishchenko92b, dluj92,
horn96}.\\

It would be interesting to investigate if comets show opposition surge when
observed in near back scattering geometry.  There have been attempts in the
past to observe the opposition surge in comets
\citep[e.g.][]{MeechJewit87,KiselevChernova81,hearn84,millis82}. Though
\citet{KiselevChernova81,hearn84,millis82} have reported enhanced 
backscattering in comets P/Ashbrook-Jackson, Bowell(1982I),  and
P/Stephan-Oterma respectively, \citet{MeechJewit87} re-analyzed the data
and found no opposition surge greater than $\sim 20\%$; instead a small
linear phase coefficient ranging $0.02~ - ~0.04$ mag deg$^{-1}$ for the
phase angle  $<30^{\circ}$ is estimated. \citet{delahodde2001} have
studied the phase function of nucleus of comet 28P/Neujmin 1 covering
$\alpha = 0.6-14.5^\circ$ and have reported a phase trend with opposition
surge.

Comet C/2007 N3 (Lulin), which was near opposition during our observing run
(i.e. February 24-28), presented an opportunity to investigate  opposition
effect. However, the comet C/2007 N3 (Lulin) happened to be close to opposition
i.e. phase angle (Sun-Comet-Earth angle $\alpha$ reaching $0^{\circ}$) at
$\sim$ UT 08:00 February 26, 2009 (local day time), by the time we could
begin observation on February 26, comet's phase angle had increased to
$1.7^{\circ}$. The comet was  bright enough to carry out high S/N  photometric
observations and in this communication we present the results based on the
photometric observations on comet C/2007 N3 (Lulin) during February 24-28,
2009 when phase angle ranged from $1.7^\circ$ to $10^\circ$.  

\section{Observations and analysis}

\begin{figure} 
\centering{
\includegraphics[width=\columnwidth]{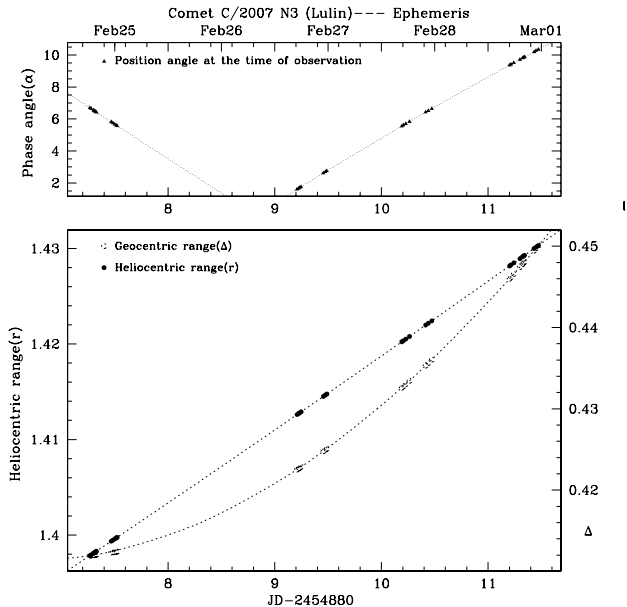}}
\caption{Upper panel: Change in phase angles ($\alpha$) with time during the
observing run. Lower panel: Change in heliocentric range($r$) and  
geocentric range ($\Delta$).  $r$ and $\Delta$ at the time
of observation are marked. Julian date in x-axis is abbreviated as JD.}
\label{Lulin_ephe}
\end{figure}

Photometric observations of comet C/2007 N3 (Lulin) were made during the
period February 24-28, 2009 with a two channel photo-polarimeter
\citep{Deshpande1985,Joshi1987}, mounted on the 1.2m telescope of Mt. Abu
IR Observatory (MIRO) operated by the Physical Research Laboratory (PRL),
Ahmedabad. The instrument is equipped with IAU's International Halley Watch
($IHW$) continuum filters (3650/80\AA (UC), 4845/65\AA (BC), 6840/90\AA (RC))
\citep{Osborn1990} and broad band filter R ($\lambda = 6400\AA;
\Delta \lambda = 1580\AA$). We have been regularly using the $IHW$ filters for
observations
of comets  \citep{Joshi1987, Sen1991, Ganesh1998, Joshi2002, Joshi2003,
ganesh2009, joshi2010} and every time before the use we check their
characteristics. Using the same filter set facilitates comparing
different comet's data observed by us.\\ 

Most of the observations were made with aperture 26"
(corresponding to projected diameter of 7770 km on February 24); however on
February 24 some observations were made with aperture 54" (projected
diameter $16150$ km). In addition to the observations with IHW
filters, we also made observations with $R$ band filter using different
apertures - 10", 20", 26" and 54" (projected diameter varying from
$\sim$ 3000 km  to 16000 km (see  Table \ref{obstab})) to study the
brightness as a function of radial distance from the comet nucleus. All
the observations were made under dark sky conditions, comet being 
more than 3 magnitudes brighter than the sky. Nonetheless, to take care of the
sky brightness, observations were made alternately centered on the photo
center of the comet and on a region of the sky more than 30 arcmin away
from the comet (along the anti-tail direction). The comet magnitudes were
calculated after subtracting sky brightness. Observed  comet magnitudes
were converted to standard system using the observations of solar type
stars HD88725, HD76151.

\begin{table*} 
\caption{Photometric observations of comet Lulin C/2007 N3. Listed
 entries are Date, JDT (=JD-2454880, where JD represent Julian date),
  right ascension(RA) 
 and declination(Dec) of the comet at the time of observation, Heliocentric 
 range($r$), Geocentric range($\Delta$), phase angle($\alpha$), aperture("), 
 projected diameter on comet, 
 filter, total integration tine(IT) and magnitude.}
\begin{tabular}{|l|l|l|l|l|l|l|l|r|r|r|r|} 
\hline
Date&JDT  & RA(H:M:S) & Dec($^{o}~'~"$) & $r(AU)$  & $\Delta(AU)$ & $\alpha^\circ$ & Ap(arcsec) &ProDia(km)& Fil & IT(sec) & mag \\
\hline
Feb 24&7.26686 & 11:7:8    & 5:42:9    &1.39784 & 0.411934  & 6.69049    &  26&  7768 &  6840&360 &   8.91$\pm$0.04  \\	
      &7.27789 & 11:6:55   & 5:43:26   &1.39792 & 0.411955  & 6.64204    &  26&  7768 &  4845&600 &   9.89$\pm$0.03  \\	
      &7.30027 & 11:6:29   & 5:46:1    &1.39809 & 0.411998  & 6.54373    &  26&  7769 &    $R$&120  &   9.05$\pm$0.03  \\	
      &7.30578 & 11:6:22   & 5:46:39   &1.39813 & 0.412009  & 6.51953    &  20&  5976 &    $R$&120  &   9.40$\pm$0.03  \\	
      &7.31339 & 11:6:14   & 5:47:32   &1.39819 & 0.412024  & 6.48614    &  13&  3885 &    $R$&150  &   9.81$\pm$0.03  \\	
      &7.32004 & 11:6:6    & 5:48:18   &1.39824 & 0.412036  & 6.45691    &  10&  2988 &    $R$&150  &  10.13$\pm$0.03  \\	
      &7.32810 & 11:5:57   & 5:49:14   &1.3983  & 0.412052  & 6.42153    &  54& 16138 &    $R$&150  &   8.35$\pm$0.03  \\	
      &7.46641 & 11:3:16   & 6:5:14    &1.39934 & 0.412319  & 5.81402    &  26&  7775 &    $R$&200  &   8.99$\pm$0.03  \\	
      &7.48224 & 11:2:58   & 6:7:4     &1.39946 & 0.41235   & 5.74453    &  26&  7776 &  6840&1000&   8.87$\pm$0.03  \\	
      &7.50100 & 11:2:36   & 6:9:14    &1.3996  & 0.412386  & 5.66213    &  26&  7776 &  4845&1000&   9.82$\pm$0.03  \\	
      &7.51507 & 11:2:20   & 6:10:52   &1.3997  & 0.412413  & 5.60032    &  54& 16152 &  6840&300 &   8.11$\pm$0.03  \\	
      &7.52227 & 11:2:11   & 6:11:42   &1.39976 & 0.412427  & 5.5687     &  54& 16153 &  4845&300 &   8.89$\pm$0.05  \\	
Feb 26&9.19308 & 10:30:2   & 9:18:10   &1.41272 & 0.422649  & 1.66324    &  26&  7970 &  6840&180 &   8.88$\pm$0.03  \\	
      &9.20926 & 10:30:18  & 9:16:39   &1.41261 & 0.422524  & 1.6026     &  26&  7968 &  6840&1400&   8.88$\pm$0.03  \\	
      &9.23272 & 10:29:52  & 9:19:6    &1.41279 & 0.422726  & 1.69999    &  26&  7971 &  4845&1200&   9.88$\pm$0.03  \\	
      &9.24727 & 10:29:35  & 9:20:36   &1.4129  & 0.422853  & 1.76037    &  26&  7974 &     $R$& 240&   9.03$\pm$0.03  \\	
      &9.45573 & 10:25:43  & 9:42:2    &1.41451 & 0.424776  & 2.62004    &  26&  8010 &     $R$& 280&   9.05$\pm$0.03  \\	
      &9.47419 & 10:25:22  & 9:43:54   &1.41466 & 0.424956  & 2.6955     &  26&  8013 &  4845& 800&   9.92$\pm$0.04  \\	
      &9.48810 & 10:25:7   & 9:45:19   &1.41476 & 0.425092  & 2.75236    &  26&  8016 &  6840& 800&   8.93$\pm$0.04  \\	
Feb 27&10.19096 &10:12:24  & 10:54:44  &1.42023 & 0.432589  & 5.5499     &  26&  8157 &     $R$& 240&   9.25$\pm$0.03  \\	
      &10.20660 &10:12:7   & 10:56:14  &1.42035 & 0.432772  & 5.61133    &  26&  8161 &  6840&1200&   9.08$\pm$0.03  \\	
      &10.22891 &10:11:43  & 10:58:22  &1.42052 & 0.433034  & 5.699      &  26&  8166 &  4845&1400&  10.08$\pm$0.03  \\	
      &10.26322 &10:11:6   & 11:1:38   &1.42079 & 0.433442  & 5.83363    &  26&  8173 &  3650&3200&  11.17$\pm$0.03  \\	
      &10.41472 &10:8:24   & 11:15:53  &1.42198 & 0.435305  & 6.42402    &  26&  8209 &     $R$& 240&   9.26$\pm$0.03  \\	
      &10.43958 &10:7:57   & 11:18:12  &1.42217 & 0.43562   & 6.52009    &  26&  8215 &  4845&2600&  10.33$\pm$0.04  \\	
      &10.47126 &10:7:22   & 11:21:16  &1.42243 & 0.436043  & 6.64736    &  26&  8222 &  6840&1400&   9.32$\pm$0.03  \\	
Feb 28&11.20142 &9:54:46   & 12:26:31  &1.42817 & 0.445986  & 9.36405    & 26 &  8410 &    $R$& 240&   9.41$\pm$0.03  \\	
      &11.21471 &9:54:33   & 12:27:39  &1.42828 & 0.44618   & 9.41268    & 26 &  8414 & 6840 &1200&   9.24$\pm$0.03  \\	
      &11.24145 &9:54:5    & 12:29:57  &1.42849 & 0.446574  & 9.51046    & 26 &  8421 & 4845 &2000&  10.27$\pm$0.03  \\	
      &11.29839 &9:53:7    & 12:34:48  &1.42894 & 0.447422  & 9.71799    & 26 &  8437 & 4845 &2000&  10.28$\pm$0.03  \\	
      &11.32226 &9:52:43   & 12:36:50  &1.42913 & 0.447781  & 9.80471    & 26 &  8444 &   $R$&400  &   9.44$\pm$0.03  \\	
      &11.33268 &9:52:32   & 12:37:43  &1.42921 & 0.447939  & 9.84251    & 20 &  6498 &   $R$&280  &   9.81$\pm$0.03  \\	
      &11.34157 &9:52:23   & 12:38:28  &1.42928 & 0.448074  & 9.87468    & 10 &  3250 &   $R$&320  &  10.58$\pm$0.03  \\	
      &11.43195 &9:50:51   & 12:46:3   &1.43    & 0.449463  & 10.2002    & 26 &  8476 &   $R$&840  &   9.51$\pm$0.03  \\	
      &11.45084 &9:50:32   & 12:47:37  &1.43015 & 0.449757  & 10.2677    & 26 &  8481 &   $R$&800  &   9.59$\pm$0.03  \\	
      &11.47224 &9:50:11   & 12:49:24  &1.43032 & 0.450092  & 10.3441    & 20 &  6529 &   $R$&400  &  10.09$\pm$0.03  \\	
\hline 
\end{tabular} 
\label{obstab} 
\end{table*}

\section{Results \& discussion}
 
\begin{figure} 
\centering{
\includegraphics[width=\columnwidth]{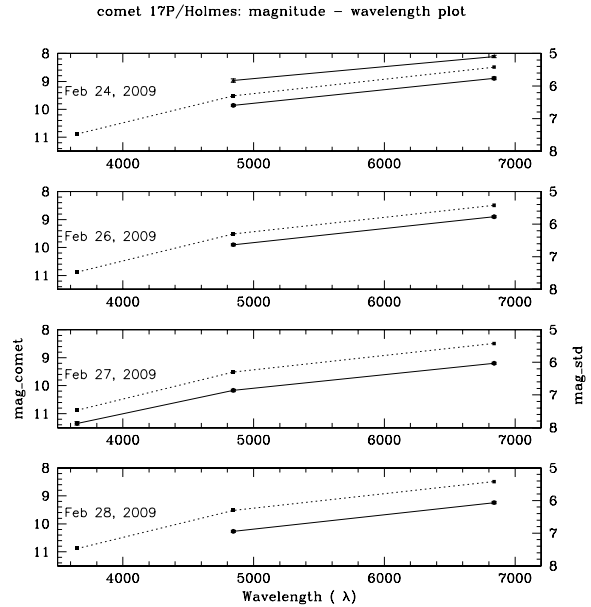}}
\caption{Spectral energy distribution as observed through 
different apertures on different dates. Observed magnitude of comet is plotted
against the mean wavelength of the filter band. Filled triangle and filled circle
represent the data observed through 54" and 26" apertures respectively and
open square is for the solar analogue (HD 76151). Error bars $\pm 1\sigma$ 
($<$ 0.05 mag), lie within the symbols used.}
\label{sed}
\end{figure}
   
The observation log with comet parameters and our results is given in
Table \ref{obstab}. The entries include date, JDT (=JD - 2454880), RA \& Dec,
Heliocentric($r$) \& Geocentric($\Delta$) distances, phase
angle($\alpha$), aperture used, projected diameter, filter, total
integration time and magnitude. The observed magnitudes reported here 
were corrected for extinction and then
converted to the standard magnitude system using the observations of the
solar analogs. To estimate instrumental magnitude, several observations of
shorter integation time (say 40 sec in R-band and 200 sec in narrow bands) were
taken and then average value of magnitude and its statistical error
were calculated. Details of the observing procedure are same as discussed in
\citet{joshi2010}. The error in magnitude listed in Table \ref{obstab} is the
total error which includes observational error, error in atmospheric extinction
and the error due to the transformation of magnitude to standard system.  \\

Figure \ref{Lulin_ephe} shows $r$, $\Delta$ and $\alpha$ at the time of
observation taken from the HORIZONS ephemeris\citep{yeoman}.
The comet was closest to the Earth on February 24, 2009 ($\Delta =$ 0.41
AU \& with 26" aperture the  projected diameter of comet is 7700 km
). The sampled area projected on the comet is not large enough to average
out the small scale inhomogeneities. There might be a possibility of
BC band being contaminated by the $C_2$ emission band lying close to it. To
address this possibility,  we looked at the spectra provided by Buil
(private communication)\footnote {see his web-site
http://www.astrosurf.com/buil/} for February 22, 2009 (the nearest date to
our observing run), and notice that $C_2$ emission band close to the BC
($4845\AA$) band partly overlaps with it. The upper limit of contamination
of the BC band by this feature is estimated to be $< 10\%$. The colour of
the comet, as discussed in section 3.1, is close to the solar colour.  This
supports the view that the contamination of comet's BC magnitude due to the
neighboring $C_2$ emission band is negligible.  Also the comet's heliocentric 
distance has not changed significantly during the observing run to affect the 
above inference.

\subsection{Spectral Energy Distribution and the Phase Curve}
\begin{figure*} 
\includegraphics[width=0.45\textwidth]{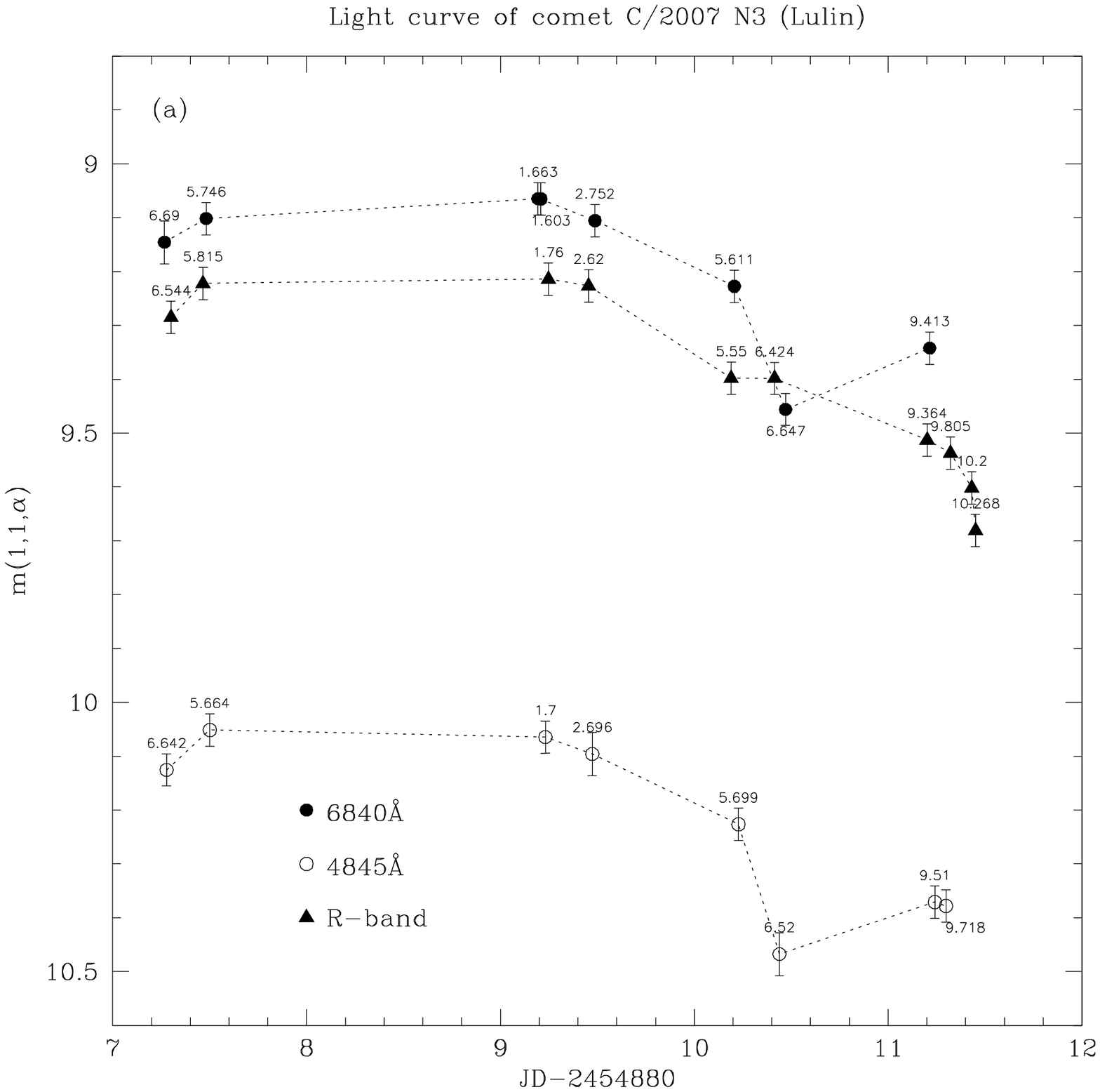}
\includegraphics[width=0.45\textwidth]{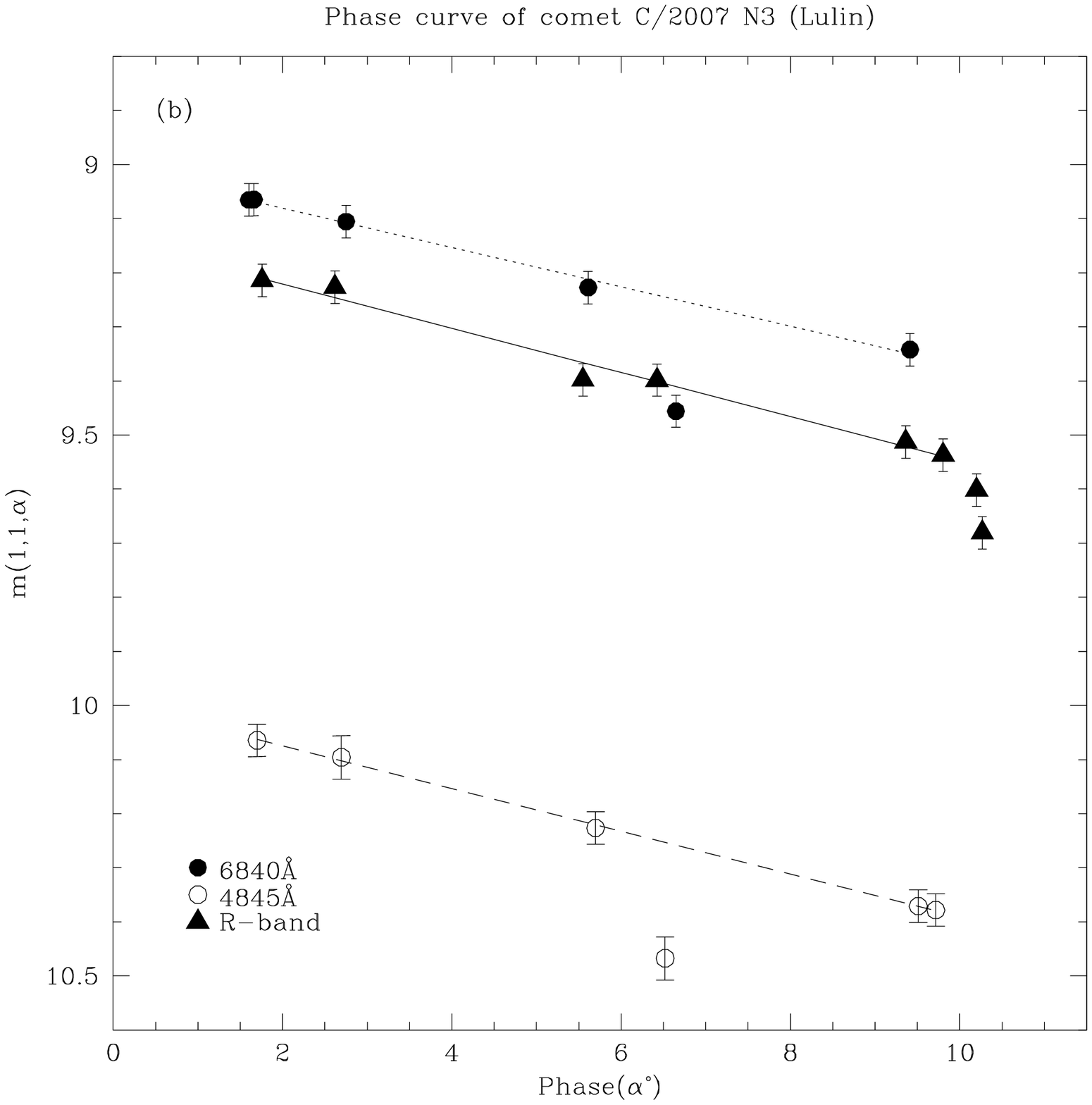} 
\caption{(a) Light curve showing variation of IHW and $R$ magnitudes  through 26" aperture normalized to $\Delta$ = 1AU, $r$=1AU)  with  time(JD - 2454880).   (b) Normalized magnitude plotted against post opposition phase angle. 
Error bars ($\pm\sigma$) are plotted.}  
\label{light_curve}
\end{figure*}
Figure \ref{sed} shows the continuum energy distribution of C/2007 N3
(Lulin). In this figure observed magnitudes are plotted
against the mean wavelength of the filter band. For comparison, energy
distribution of solar analogue star HD76151, which was observed during the
observing run, is  also plotted.  It is seen that
during the observing run the comet colour is similar to that of the solar
colour. On February 24 the comet is observed through two apertures: 26"
and 54" (projected diameter 7700 km and 16300 km respectively). There is
indication that comet colour through the larger aperture  is slightly
bluer. This could be due to the disintegration of larger grains into
smaller grains as they move out resulting in higher population of
sub-micron size grains in the outer coma. Observations on all the dates
with 26" aperture show comet colour similar to solar colour.\\

In the following we discuss the light curve and the phase curve of the comet.
For this purpose, magnitudes, denoted as $m(1,1,\alpha)$, are
referred to at $\Delta$ and $r$ equal to 1AU by subtracting the term
2.5Log($\Delta ~r^2$) from the observed magnitudes.
The light curves (LCs i.e. m(1,1,$\alpha$) vs time(JD)) of C/2007 N3 (Lulin)
are plotted in Figure \ref{light_curve}(a) and the phase angles at the time
of observation are marked. LCs in different filter bands are annotated in
the figure. It is seen that comet gets brighter as $\alpha$ decreases. During the
pre-opposition phase we have observations only at two phase angles - 6.69
and $5.75^\circ$, with the comet being brighter at lower
phase angle. The phase curve is better covered during the
post-opposition phase, the minimum phase angle, at which observations are
made, being $\sim 1.7^\circ$. The comet was not observable when it was
close to zero phase (local day time). During post-opposition phase, a
clear increase in brightness with decreasing $\alpha$ is observed. This is
consistent in all the filter bands. To look at the change in brightness
with phase, we have plotted  m(1,1,$\alpha$) vs $\alpha$ in Figure
\ref{light_curve}(b).  No opposition surge is noticed for the observed
range in $\alpha$, only some linear increase of brightness with
decreasing $\alpha$ is detected.
We checked for the linearity
in phase curve by plotting the flux against the phase angle and found it to
be linear in all the bands. For discussion purpose in this paper we have
adopted the phase curve (magnitude vs phase) i.e. Figure \ref{light_curve}(b). Though
we notice a linear brightness increase with decreasing $\alpha$ for
$1.7^{\circ} <\alpha < 10^{\circ}$, we do not have observations for
$\alpha < 1.7^{\circ}$ to comment on opposition surge (i.e. 
any non-linear increase in
brightness ) as $\alpha$ approaches zero degree.\\ 

Phase curve shows one interesting feature - at phase angle near
$6.5^\circ$ the brightness shows a decrease in BC and RC bands. As
there are no observations (including the R band) between phase angle 6.4
and $9.4^\circ$, it is difficult to comment on the actual trend. 
The phase curve in $R$ band shows a sudden decrease in brightness at
phase angle $> 10^\circ$ but unfortunately we do not have data in BC and
RC in this  range to make a comparison.  As discussed earlier, the
sampled area projected on the comet is not large enough to average out the
small scale inhomogeneities which might cause fluctuations in the brightness. 
However, with the present data it is difficult to quantify this.

We have estimated the phase coefficient (i.e. slope of the phase curves)
ignoring the dip, discussed in the previous paragraph. The estimated slopes
in 4845\AA, 6840\AA and R-bands are respectively $0.036\pm 0.002$,~ $0.040\pm
0.001$ and $0.041\pm 0.003$ mag deg$^{-1}$. Within the estimated error, the
phase coefficients in all the filter bands might be taken same
i.e. the phase coefficient is independent of the wavelength. The shadow 
hiding model can explain this kind of brightening of the comet with
decreasing phase angle. \citet{MeechJewit87} have reported similar results
on some other comets studied by earlier researchers(see section 1). However, 
the phase coefficient in the present case is slightly higher than the values
reported by \citet{MeechJewit87}. Though \citet{delahodde2001} have reported
a phase trend with opposition surge ($\alpha = 0.6-14.5^\circ$) in case of
the nucleus of comet 28P/Neujmin 1, it is very likely that comet nucleus surface
behaves differently (similar to asteroids) than the  dust particles in the
coma.

\subsection{Radial Profile of the brightness} 
\begin{figure} 
\centering{
\includegraphics[width=\columnwidth]{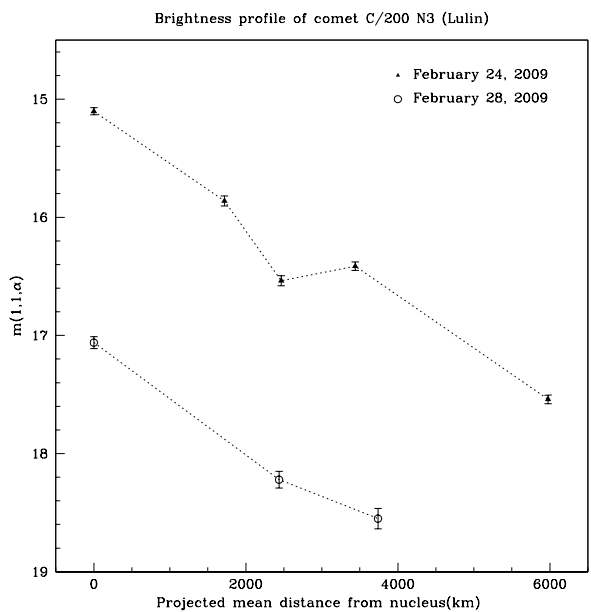}}
\caption{Radial brightness variation as observed  
on February 24 and 28, 2009. Error bars are $\pm \sigma$.}
\label{rad_brightness}
\end{figure}

As mentioned earlier in section 2, we have made observations in $R$-band
through various apertures (10'', 20'', 30'' \& 54'')
 to study behaviour of the comet brightness with distance
from the photo-center. Radial profile of brightness in $R$-band is plotted
in Figure \ref{rad_brightness} for February 24 and 28. The procedure of
getting the radial distribution of the brightness is the same as described in
our earlier papers \citep {ganesh2009, joshi2010}. On February 24 brightness
drops at a distance of $\sim $ 2000km.  As the sampled area on comet is
not large, the possibility of observing inhomogeneous projected area
can not be ruled out.  Later, on February 28 the brightness distribution is
smooth apart from a slight depression beyond 2000 km.

\section{Conclusions}

The following conclusions are drawn from the present study on 
Comet C/2007 N3(Lulin).  

\begin{enumerate}

\item Colour of the comet is similar to the solar colour indicating the grain
size larger than $0.1 \mu m$.

\item Significant brightness enhancement with decreasing phase angle is
observed. The phase coefficient, $\beta$, is found to be $0.040\pm 0.001$ mag
deg$^{-1}$ which is independent of the wavelength. These findings 
support shadow hiding model to explain the increase in brightness with decreasing 
phase angle.  

\item A dip of $\sim$ 0.20 mag in the brightness at the phase angle $\sim
6.5^\circ$ is observed in the continuum narrow bands which might be due to
presence of inhomogeneities in the coma.

\item Brightness drops smoothly radially outward apart from a slight
depression near 2000km, especially on February 24.

\end{enumerate}

\section{Acknowledgments}

The authors thank the referees for their very constructive and critical 
comments which has improved the content of this paper.  
We thank Christian Buil, Castanet-Tolosan Observatory (France), for 
allowing us to use the spectrum of C/2007 N3 (Lulin) taken by him. 
The work reported here is supported by the Department of Space, 
Government of India.

\bibliographystyle{mn2e}
\bibliography{comet}
\bsp
\label{lastpage}
\end{document}